\documentclass[letterpaper, 10pt, conference]{ieeeconf}      

\IEEEoverridecommandlockouts                              

\title{\LARGE \bf
	Event-based analysis of neuromorphic feedback systems
}

\author{Alessandro Cecconi$^{1,2}$, Michelangelo Bin$^{2}$, Lorenzo Marconi$^{2}$, Rodolphe Sepulchre$^{1,3}$
	\thanks{$^{1}$Alessandro Cecconi and Rodolphe Sepulchre are with STADIUS, KU Leuven, Belgium
		{\tt\small \{alessandro.cecconi, rodolphe.sepulchre\}@kuleuven.be}}%
	\thanks{$^{2}$Alessandro Cecconi, Michelangelo Bin and Lorenzo Marconi are with DEI, University of Bologna, Italy
		{\tt\small \{a.cecconi, michelangelo.bin, lorenzo.marconi\}@unibo.it}}%
\thanks{$^{3}$Rodolphe Sepulchre is with the Department of Engineering, Cambridge, UK.
This work was supported by the ERC Advanced Grant SpikyControl, n.~101054323.}
}

\usepackage{hyperref}
\usepackage{amsmath}

\usepackage{amsthm}
\usepackage{amssymb}
\usepackage{amsfonts}
\usepackage{mathtools}
\usepackage{xparse}
\usepackage{environ}
\usepackage{physics}

\usepackage{graphicx} 
\usepackage{float}
\usepackage{pgfplots}
\usepackage{subcaption}
\usepackage{color}
\usepackage{wasysym}
\usepackage{nicematrix}
\usepackage{ipsum}
\pgfplotsset{compat=1.18}

\usepackage{tikz}
\usepackage{circuitikz}
\usetikzlibrary{shapes,arrows,positioning,calc,fit,arrows.meta}

\tikzset{
output/.style ={coordinate}
}

\tikzstyle{neuron} = [
circle,
minimum size=0.5cm,
color=black,
fill=black!5,
draw=black,
inner sep=2pt
]

\tikzstyle{sum} = [
draw,
fill=white!20,
circle,
inner sep = 0pt,
minimum width=0.2cm,
anchor=center
]

\tikzstyle{block} = [
draw,
color=black,
fill=black!5,
rectangle,
rounded corners=0pt,
minimum height=0.75cm,
minimum width=2.75cm,
line width=0.5pt,
anchor = west,
align = center,
text=black,
inner sep=2pt
]

\tikzstyle{lines} = [
black,
line width=0.5pt,
->
]

\theoremstyle{plain}
\newtheorem{proposition}{Proposition}

\newtheorem{theorem}{Theorem}
\newtheorem{corollary}{Corollary}
\theoremstyle{definition}
\newtheorem{definition}{Definition}

\theoremstyle{remark}

\newcommand{\set}[1]{\left \{ {#1} \right \}}

\newcommand{\cW}{\mathcal{W}}
\newcommand{\cF}{\mathcal{F}}

\newcommand{\Z}{\mathbb Z}

\newcommand{\cD}{\mathcal D}

\newcommand{\cR}{\mathcal R}

\newcommand{\cT}{\mathcal T}

\newcommand{\lo}{\text{low}}
\newcommand{\hi}{\text{high}}

\newcommand{\st}{\mid}

\newcommand{\one}{\mathbf{1}}       

\begin{document}
	\setlength{\parskip}{0pt}
	\setlength{\parindent}{1em}

	\maketitle
	\thispagestyle{empty}
	\pagestyle{empty}
	
	\begin{abstract}
		This article investigates the existence and stability of periodic behavior arising in the feedback interconnection of neuromorphic systems, which evolve according to continuous dynamics but interact only through discrete events. Exploiting the event-driven nature of this interaction, each system is modeled by an operator that maps the history of events to the delay of its response. Building on the fading-memory properties of these operators, the article develops a small-gain result establishing the existence, uniqueness, and robustness of periodic one-to-one rhythms.
	\end{abstract}

	\section{Introduction}\label{sec:intro}
	
	Rhythmic behavior is pervasive in biological, physical, and engineered
	systems, and is central to neuromorphic feedback
	control~\cite{buzsaki2006rhythms,nayfeh2024nonlinear,sepulchre2026regulation}.
	From a dynamical systems viewpoint, such rhythms correspond to
	limit cycles of the nonlinear closed-loop system. Establishing
	their existence and stability is a classical problem, but an analysis
	performed in the state space is out of reach in all but simple cases
	\cite{guckenheimer_nonlinear_1983,Khalil2002}.
	Approximate methods such as the describing function have proved effective in
	reducing the analysis to a lower-dimensional problem, at the price of
	neglecting the higher harmonics of the nonlinear response
	\cite{vander1968multiple}.
	
	The problem is of particular relevance in neuromorphic control. For instance,
	\cite{schmetterling2024neuromorphic} investigates the control of a pendulum
	by an excitable circuit, in a configuration where the interaction between
	the plant and the controller is purely event-based.
	The feedback analysis of the resulting closed-loop system, however, remains
	an open question, due to the intricacy of analyzing a limit cycle
	in a high-dimensional nonlinear state-space model.
	The objective of the present paper is to address this question from a
	\emph{neuromorphic} viewpoint, focusing on the discrete events that
	characterize the closed-loop interaction rather than on the underlying
	continuous dynamics~\cite{sepulchre2022spiking}. Indeed,
	between interaction events, the systems evolve independently, while their
	interconnection is active only at event times.
	This aligns neuromorphic systems with the event-based control paradigm
	envisioned by Astr\"om~\cite{astrom2008event}, in which each subsystem
	evolves mostly in open loop, and feedback is active only at events times.
	
	This paper exploits such a separation between continuous dynamics and
	event-based interaction to formulate the feedback analysis directly
	in the event domain. Each system is described by an \emph{event
		operator}, which maps sequences of incoming events to the delays
	of the output events they trigger. Closing the loop then amounts to
	interconnecting the event operators of the two subsystems, yielding a
	fixed-point equation directly in terms of the exchanged response delays.
	A small-gain theorem is then established in the event domain,
	guaranteeing the existence, uniqueness, and robustness to timing
	perturbations of periodic one-to-one rhythms.
	
	These rhythms have a particularly simple interpretation. The events
	of the two subsystems alternate, each event triggering exactly one
	response from the other. Each subsystem is therefore periodically
	forced, and its steady-state response is characterized by a single
	delay determined by the forcing period. The closed-loop period is
	thus determined by a scalar balance between the periodic responses
	of the two subsystems, in the spirit of the event-based describing
	function recently introduced in~\cite{wroblowski2026describing}.
	
	This event description is natural for excitable systems, which remain at
	rest in the absence of stimulation and respond to an incoming event
	with a transient excursion before relaxing back to
	rest~\cite{izhikevich_dynamical_2006}.
	The framework is nevertheless more general, as events can also be
	extracted from continuous trajectories through suitable \emph{event readouts},
	as exemplified by event-based sensors
	\cite{liu2010neuromorphic,gallego2020event}.
	
	A key property motivating this description is contraction of the
	event-driven dynamics~\cite{lohmiller_contraction_1998,bullo_contraction_2024}.
	Recent results have shown that excitable systems exhibit contractive
	behavior under suitable conditions on the temporal spacing of incoming
	events~\cite{cecconi2025contraction,bin2025reliability}.
	In particular, when successive events are sufficiently spaced in time,
	the transitions between events can become contractive, so that
	trajectories driven by the same input progressively lose their
	dependence on initial conditions.
	While contraction is a system property, the related
	input–output notion of \emph{fading memory} describes how the
	influence of past inputs on the current output diminishes over time
	\cite{boyd1985fading}.
	Recent work has provided a state-space characterization of fading
	memory~\cite{bainier2026state}, while contraction, uniform over a compact
	set of inputs, has been shown to imply fading memory whenever the
	output map is continuous~\cite{matthews1993approximating}.
	These results motivate the use of event operators with fading memory
	as the basis for the feedback analysis developed in this paper.
	
	\emph{Notation.}
	$\Z$ denotes the integers and $\mathbb N\coloneqq\{1,2,\dots\}$. For
	$\mathcal A\subseteq\mathbb R$, $\mathcal A^{\Z}$ and $\mathcal A^{\mathbb N}$ are
	the sets of sequences with values in $\mathcal A$ indexed by $\Z$ and
	$\mathbb N$, and we let $\|x\|_\infty\coloneqq\sup_k|x_k|$. Sequences in
	$\mathcal A^{\Z}$ are the signals on which the event operators act, and are
	written in bold, such as $\mathbf T=\{T_k\}_{k\in\Z}$, while sequences in
	$\mathcal A^{\mathbb N}$ are input histories, such as
	$v=\{v_j\}_{j\in\mathbb N}$. The shift operator is defined as $(\sigma\mathbf T)_k\coloneqq T_{k+1}$,
	and $\one$ will denote the constant sequence of ones. For a function $z$ of time,
	$z(t^-)$ and $z(t^+)$ denote its left and right limits at $t$, respectively. The proofs of all results are collected in the Appendix.
	
	\section{Motivating question}\label{sec:motivating}
	
\begin{figure}[t]
		\centering
		\includegraphics[width=0.45\textwidth]
		{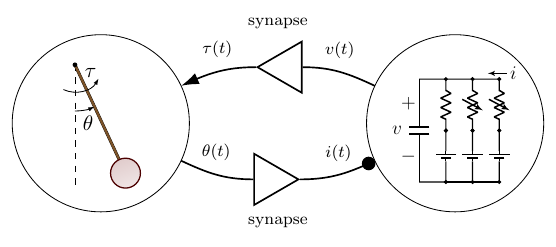}
		\caption{Feedback interconnection of an excitable neuron and a damped
			pendulum. An excitatory synapse converts neuron voltage $v(t)$ into a torque signal $\tau(t)$ driving the pendulum, while an inhibitory synapse converts the angle signal $\theta(t)$ into a current $i(t)$ hyperpolarizing the neuron.}
		\label{fig:interconnection}
	\end{figure}

	Consider the feedback interconnection of an excitable neuron and a damped
	pendulum illustrated in Fig.~\ref{fig:interconnection}. The two systems
	interact through an excitatory and an inhibitory synapse: the former
	converts the membrane voltage of the neuron into a force acting on the
	pendulum, while the latter converts the pendulum angle into a current
	injected into the neuron. The resulting system is a standard feedback loop described by a set of
	nonlinear differential equations. Both systems are non-oscillatory in
	isolation: in the absence of interaction, the neuron converges to its
	resting equilibrium, while the damped pendulum relaxes towards its stable
	equilibrium. Nevertheless, their interconnection can sustain a limit-cycle
	oscillation~\cite{schmetterling2024neuromorphic}.
	
	Analyzing the stability of the resulting oscillation directly in
	the trajectory space is a classical but challenging problem. A
	conductance-based neuron typically has four state variables, the
	pendulum has two, and each synapse introduces at least one additional
	state. Even this simple interconnection therefore results in a
	nonlinear dynamical system of dimension eight or higher, making the
	analytical characterization of its periodic solutions and their
	stability generally intractable.
	
\begin{figure}[t]
		\centering
		\includegraphics[width=0.45\textwidth]
		{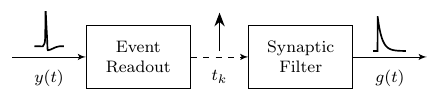}
		\caption{Event model of a synapse. An event readout extracts the event
			times $t_k$ from a presynaptic signal $y(t)$, while a synaptic
			filter, whose state is reset at each $t_k$, generates a
			conductance $g(t)$ that drives the postsynaptic system.
			Details of both models are given in the Appendix.}
		\label{fig:readout-synapse}
	\end{figure}

	Our objective is to simplify the analysis by adopting a neuromorphic perspective, that is, by exploiting the event-driven nature of synaptic interactions. A presynaptic spike triggers a biophysical reaction that
	transiently modifies the synaptic conductance, thereby enabling a continuous-time synaptic input current~\cite[Ch.~7]{ErmentroutTerman2010}. To capture this
	mechanism explicitly, we model a synapse as the cascade of two
	distinct elements, as illustrated in Fig.~\ref{fig:readout-synapse}. The first
	is an \emph{event readout}, which extracts a discrete event from the
	continuous trajectory of the presynaptic signal. The second is a \emph{synaptic filter}, whose state is
	reset at each event and which generates the physical postsynaptic signal. This decomposition introduces a well-defined
	interface between the two elements, enabling the mathematical representation of each \emph{event} by its time of occurrence. Although the signals on either side of
	this interface are continuous, the information transmitted
	between them is entirely encoded in the event times.
	
	This modeling choice augments the continuous-time physical representation of the interaction with a discrete event representation. From a neuromorphic perspective, each system of the feedback interconnection is now regarded as a causal shift-invariant operator mapping a sequence of input events to a sequence of output events.
	
	This leads to the central question of the paper: can the stability
	of the feedback interconnection be analyzed directly in event space,
	without studying the potentially intractable closed-loop dynamics
	in the trajectory space? To answer this question, we cut the feedback
	loop at the event interfaces and characterize each subsystem through
	an input-output operator acting on event sequences. We then develop
	a small-gain theorem for the interconnection of these operators,
	establishing tractable conditions for closed-loop stability under
	suitable assumptions on the underlying physical dynamics.
	
	\section{The event operator}\label{sec:model}
	
\begin{figure}[t]
		\centering
		\includegraphics[width=0.9\columnwidth]{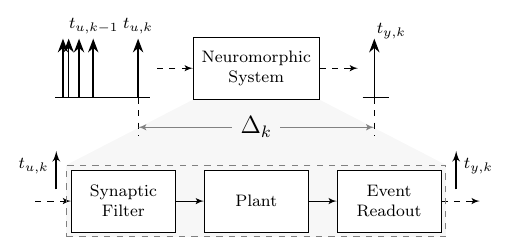}
		\caption{Block representation of a neuromorphic system. \emph{Top}: its event
			operator, in which each input event $t_{u,k}$
			produces an output event $t_{y,k}$ after a delay $\Delta_k$
			that depends on the history of input events.
			\emph{Bottom}: the underlying cascade, in which input
			events activate a synaptic filter, the filter drives a continuous-time
			plant, and an event readout extracts the output events.
			Dashed arrows represent events, solid arrows continuous-time signals.}
		\label{fig:block-diagram}
	\end{figure}

	The neuromorphic system considered in this paper is a physical system described by differential equations, consisting of a synaptic filter, a plant, and an event readout connected in cascade, as illustrated in Fig.~\ref{fig:block-diagram}. However, owing to the modeling choices introduced above, its input-output behavior can be characterized entirely in terms of input and output events, without explicitly referring to the continuous-time dynamics. We can therefore regard the system as a mapping from input event sequences to output event sequences, and we call this mapping the \emph{event operator} of the system. Since it describes a physical system, the event operator is required to satisfy two fundamental properties: causality and time invariance.
	
	Causality means that an output event can only depend on past input events. Denoting by $t_{u,k}$ the last input event
	preceding the output event $t_{y,k}$, there exists a map $F$ such that
	\begin{equation}\label{eq:causal}
		t_{y,k}=F(t_{u,k},t_{u,k-1},t_{u,k-2},\dots),
	\end{equation}
	which may depend on the entire past of the input. Time invariance
	requires that shifting all input events by a common offset $c$ shifts
	the output event by the same amount, namely
	\begin{equation}\label{eq:invariance}
		F(t_{u,k}+c,t_{u,k-1}+c,\dots)
		=F(t_{u,k},t_{u,k-1},\dots)+c,
	\end{equation}
	for all $c\in\mathbb R$. Choosing $c=-t_{u,k}$ sets the last input
	event to the origin. Combining~\eqref{eq:causal} and
	\eqref{eq:invariance} then yields
	\[
	t_{y,k}-t_{u,k}
	=F(0,t_{u,k-1}-t_{u,k},t_{u,k-2}-t_{u,k},\dots).
	\]
	Hence, the delay $\Delta_k\coloneqq t_{y,k}-t_{u,k}$ depends only
	on the relative timing of past input events, rather than on their
	absolute occurrence times. These relative times can be expressed
	as sums of inter-event intervals, defined by
	$T_k\coloneqq t_{u,k+1}-t_{u,k}$, since
	\[
	t_{u,k-j}-t_{u,k}=-(T_{k-1}+\dots+T_{k-j}), \quad k\in\Z,\ j\in\mathbb N.
	\]
	Consequently, the delay admits the representation
	\begin{equation}\label{eq:functional}
		\Delta_k=\psi(T_{k-1},T_{k-2},\dots), \quad k \in \Z,
	\end{equation}
	where the functional $\psi$ is defined as
	\begin{equation}
		\psi(v_1,v_2,\dots)
		\coloneqq F(0,-v_1,-v_1-v_2,\dots).
	\end{equation}
	We call $\psi$ the \emph{memory functional} of the neuromorphic system,
	as it describes how the delay of an output event depends on the
	history of input inter-event intervals.
	
	The common index $k$ in~\eqref{eq:causal} assumes that each input event has its own
	output event, which is not true in general: an input event may be followed by several
	output events, or by none. We therefore restrict the analysis to the following regime.
	
	\begin{definition}[One-to-one entrainment]\label{def:entrainment}
		A strictly increasing input sequence $\{t_{u,k}\}_{k\in\Z}$ entrains the system
		one-to-one if it produces an output sequence $\{t_{y,k}\}_{k\in\Z}$ with exactly one
		output event between any two consecutive input events, that is,
		$t_{u,k}<t_{y,k}<t_{u,k+1}$ for all $k\in\Z$.
	\end{definition}
	
	One-to-one entrainment cannot hold for arbitrary input sequences. When two input events
	are too close, the second one can arrive before the first has produced its output event.
	When they are too far apart, a single input event may produce several output events, or
	none. We therefore restrict the inter-event intervals to a compact set
	$\cT\coloneqq[T_{\min},T_{\max}]$, and we assume throughout the paper that every input
	sequence with $T_k\in\cT$ for all $k$ entrains the system one-to-one. The lower bound
	$T_{\min}$ is a minimum dwell time, which quantifies the sparsity of the events. The upper bound $T_{\max}$ will not be constrained by the subsequent analysis, its presence only excludes excessively long intervals without events.
	
	The previous assumption also restricts the delays that the system can
	generate. By Definition~\ref{def:entrainment}, each output event precedes the
	next input event, that is, $0<\Delta_k<T_k$. However, the delay $\Delta_k$
	does not depend on $T_k$ by~\eqref{eq:functional}: the system cannot know when
	the next input event will arrive. The output event must therefore occur before
	the next input event even when this arrives as early as possible, which gives
	$\Delta_k<T_{\min}$. We strengthen this bound by requiring that the delays lie
	in a compact interval $\cD\subset(0,T_{\min})$, so that they stay uniformly
	away from the boundaries. We can now give the definition used in the
	remainder of the paper.
	
	\begin{definition}[Event operator]\label{def:functional}
		Let $\cT=[T_{\min},T_{\max}]$ with $T_{\min}>0$, and let $\cD\subset(0,T_{\min})$ be
		a compact interval. A \emph{memory functional} is a map
		$\psi\colon\cT^{\mathbb N}\to\cD$, and the associated \emph{event operator} is
		\begin{equation}\label{eq:operator}
			\Psi\colon\cT^{\Z}\to\cD^{\Z}, \
			(\Psi\mathbf T)_k\coloneqq\psi(T_{k-1},T_{k-2},\dots),\ k\in\Z.
		\end{equation}
	\end{definition}
	
	By construction, $\Psi$ is strictly causal, since $\Delta_k$ does not depend on $T_k$,
	and shift invariant, that is $\Psi\sigma=\sigma\Psi$.
	
	The simplest, yet fundamental, input is a periodic train. For $\mathbf T=T\one$, $T\in\cT$, the delay
	is constant, namely $\Delta_k=\Delta(T)\coloneqq\psi(T,T,\dots)$, so that the output events
	are periodic with the same period. We call $T\mapsto\Delta(T)$ the \emph{event describing function} of the system, in analogy with the classical describing function analysis \cite{wroblowski2026describing}, and it can be numerically estimated by forcing the system in open loop.
	
	\subsection{Fading memory properties of event operators}
	
	Fading memory is a stability property of input-output operators that captures how the influence of past inputs progressively vanishes with time. In particular, two input histories that agree over a sufficiently recent time window should produce similar current outputs, even if they differ substantially in the more distant past. Boyd and Chua~\cite{boyd1985fading} formalized this notion by introducing a weighted norm on input histories, in which contributions from the remote past are progressively discounted. In our framework, an input history is represented by a sequence $v=(v_1,v_2,\dots)$, where $v_j$ denotes the interval of age $j$, namely, $v_j=T_{k-j}$ in~\eqref{eq:functional}. Accordingly, the weighting of past inputs translates into a sequence of weights indexed by the age of each interval.
	
	\begin{definition}[Fading memory]\label{def:fading}
		A \emph{memory kernel} is a nonincreasing sequence
		$w=\{w_j\}_{j\in\mathbb N}\subset(0,1]$ with
		$\lim_{j\to\infty}w_j=0$.
		The functional \eqref{eq:functional} has \emph{fading memory}
		with kernel $w$ if it is continuous on $\cT^{\mathbb N}$ with respect to
		the norm $\|v\|_w\coloneqq\sup_{j\ge1}w_j|v_j|$.
	\end{definition}
	Two histories that differ only in their remote past are close in this norm and therefore
	produce similar output delays. The event operators considered in this paper are assumed to have fading memory in the following quantitative sense: there exist weights $\rho_j\ge0$, with $\sum_{j\ge1}\rho_j<\infty$, such that, for all $v,\tilde v\in\cT^{\mathbb N}$,
	\begin{equation}\label{eq:fading}
		|\psi(v)-\psi(\tilde v)|\le\sum_{j\ge1}\rho_j|v_j-\tilde v_j|.
	\end{equation}
	The weight $\rho_j$ bounds the influence of the interval of age $j$. In
	particular, two histories that agree on their last $J$ intervals produce
	delays that differ by at most $(T_{\max}-T_{\min})\sum_{j>J}\rho_j$,
	whatever their remote past, and this bound vanishes as $J$ grows. This
	uniform forgetting gives fading memory regardless of the kernel, as shown in the following proposition.
	
	\begin{proposition}\label{prop:fading}
		If the functional~\eqref{eq:functional} satisfies~\eqref{eq:fading}, then
		it is uniformly continuous on $\cT^{\mathbb N}$ with respect to
		$\|\cdot\|_w$, for every memory kernel $w$. In particular, it has fading
		memory with every memory kernel, in the sense of
		Definition~\ref{def:fading}.
	\end{proposition}
	
	Fading memory describes how the delay depends on the past of the input. However, in a feedback loop, we must also know how much a timing perturbation is amplified, since it travels around the loop and can affect all the intervals at once. This amplification is measured by the \emph{incremental gain} of the operator~\cite{zames1966input}, the smallest $\gamma$ such that \[\|\Psi\mathbf T-\Psi\tilde{\mathbf T}\|_\infty\le\gamma\|\mathbf T-\tilde{\mathbf T}\|_\infty\] for all $\mathbf T,\tilde{\mathbf T}\in\cT^{\Z}$. Bounding each interval difference in~\eqref{eq:fading} by the supremum norm gives $\gamma\le\sum_{j\ge1}\rho_j<\infty$, so our fading memory assumption ensures a finite incremental gain. This incremental gain will be the only constant of each system entering the analysis of Section~\ref{sec:feedback}. Note, moreover, that since constant sequences are admissible inputs, the event describing function is also $\gamma$-Lipschitz.
	\section{Shaping the gain of the event operator}\label{sec:gain}
	
	We now return to the cascade of Fig.~\ref{fig:block-diagram} and ask under which
	conditions it defines an event operator with the properties required in
	Section~\ref{sec:model}, namely one-to-one entrainment and fading memory. We show that the determining factor is the sparsity of events:
	sparsely triggered systems define event operators with fading memory, and their
	gain decreases as the events become sparser.
	
	To illustrate the idea, we take as an example a stable first-order
	lag in series with a threshold-crossing readout, and we analyze its event operator. Each
	input event makes the state of the filter jump, which can be equivalently
	represented by a Dirac pulse at the input, so that we obtain
	\begin{equation}\label{eq:lti}
		\dot y=-\frac{y}{\tau}+b\sum_{k\in\Z}\delta(t-t_{u,k}),
	\end{equation}
	where $b>0$ is the amplitude of the jump, $\tau>0$ is the system time constant, and
	$\delta$ denotes the Dirac distribution. An output event is produced every time $y(t)$
	crosses a threshold $y_{\rm th}\in(0,b)$ from above. Just after the input $t_{u,k}$,
	the state collects the response to this input and the residuals of all the previous
	ones, obtaining
	\[
	y(t_{u,k}^+)=b\sum_{j\ge0}e^{-a_j/\tau},
	\]
	where $a_0\coloneqq0$ and $a_j\coloneqq T_{k-1}+\dots+T_{k-j}$ is the age of the $j$-th
	previous input. The output then reaches the threshold after a delay $\Delta_k$ such that
	\[
	y(t_{u,k}^+)e^{-\Delta_k/\tau}=y_{\rm th}.
	\]
	Writing the ages in terms of a history $v$, $a_j=v_1+\dots+v_j$, we obtain the memory
	functional
	\begin{equation}\label{eq:lag-functional}
		\psi(v)=\tau\ln\left(
		\frac{b}{y_{\rm th}}\sum_{j\ge0}e^{-a_j/\tau}
		\right).
	\end{equation}
	The residuals represent the \emph{memory} of the system: they lengthen the delay with
	respect to that of an isolated response, $\tau\ln(b/y_{\rm th})$, and they vanish as the
	time between events increases. For a periodic train we have $a_j=jT$, and we can
	explicitly compute its event describing function as
	\begin{equation}\label{eq:lag-periodic}
		\Delta(T)=\tau\ln\frac{b/y_{\rm th}}{1-e^{-T/\tau}}.
	\end{equation}
	
	The operator is well defined on $\cT$ if the output falls below the threshold before the
	next input arrives, which holds when $T_{\min}>\tau\ln(1+b/y_{\rm th})$, as shown in the
	Appendix. In other words, the input period must exceed the system time constant, up to a
	logarithmic factor, so that the filter has enough time to relax between events.
	
	We now ask how much the operator amplifies timing differences, and how this
	depends on the time between events. We start from the periodic train of period
	$T$, whose history is $v=(T,T,\dots)$, and we perturb its intervals by
	$\mathrm dT_k$. The partial derivatives of~\eqref{eq:lag-functional} at this
	history are $\partial\psi/\partial v_j=-e^{-jT/\tau}$, so that, to first order, we have
	\begin{equation}\label{eq:lag-sensitivity}
		\mathrm d\Delta_k=-\sum_{j\ge1}e^{-jT/\tau}\,\mathrm dT_{k-j}.
	\end{equation}
	An interval therefore acts on the $j$-th following delay with weight
	$e^{-jT/\tau}$, the fraction of $y(t_{u,k}^+)$ carried by the inputs at least
	$j$ events old. For perturbations of the intervals of unit size, the change of
	a delay is largest when each $\mathrm dT_{k-j}$ has the sign of its weight. We
	call this largest amplification the \emph{local gain}, and it is the sum of the
	absolute values of the weights, namely
	\begin{equation}\label{eq:lag-gain}
		\gamma_{\rm loc}(T)=\sum_{j\ge1}e^{-jT/\tau}=\frac{1}{e^{T/\tau}-1}.
	\end{equation}
	Since all the weights have the same sign, the worst case is to perturb all the
	intervals equally, that is, to change the period of the train, so that
	$\gamma_{\rm loc}(T)=|D\Delta(T)|$.
	
\begin{figure}[ht]
		\centering
		\includegraphics[width=0.425\textwidth]{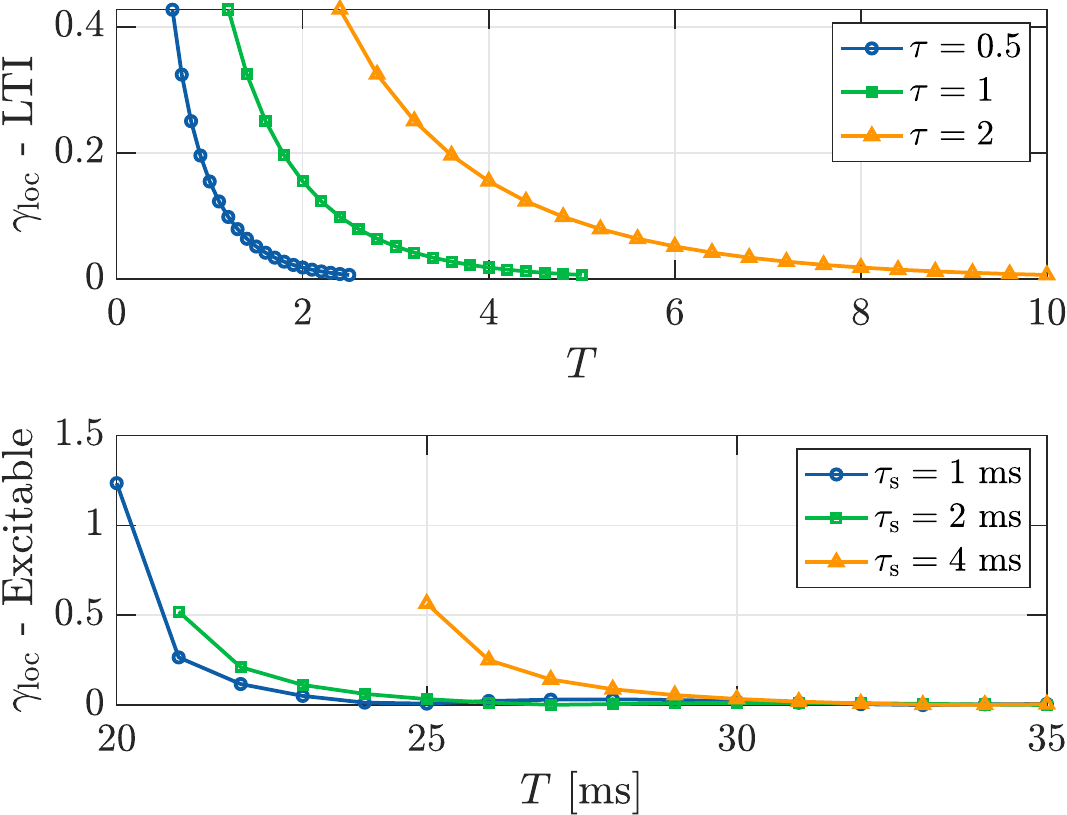}
		\caption{Local gain of the event operator around a periodic train of period $T$. \emph{Top}:
			First-order lag with $b/y_{\rm th}=2$, for three system time constants. \emph{Bottom}:
			Conductance-based neuron forced through an inhibitory synapse for three different
			synaptic time constants.}
		\label{fig:event-gain}
	\end{figure}

	The local gain depends only on the ratio $T/\tau$ and decays exponentially with it:
	when the time between events is much larger than the time constant, the linear system
	relaxes completely between events and the delay no longer depends on the past.
	Fig.~\ref{fig:event-gain}(Top) shows this decay, measured by lengthening one interval
	of a train of period $T$ and summing the absolute changes of the following delays.
	The local gain covers only infinitesimal perturbations of one periodic train, and in
	general it only bounds from below the incremental gain $\gamma$, which covers arbitrary
	differences between arbitrary sequences. For this system, however, the two are linked: as shown
	in the Appendix, \eqref{eq:fading} holds with $\rho_j=e^{-jT_{\min}/\tau}$, so the
	operator has fading memory, and $\gamma=\gamma_{\rm loc}(T_{\min})$, attained at the
	densest admissible train. The incremental gain therefore also decreases as the time
	between events increases.
	
	The same idea applies to a neuromorphic system made of a synaptic filter, an excitable
	neuron, and an event readout. The role of the system time constant is played here by the
	time the neuron needs to recover from the excursion triggered by an input event: a
	sufficiently spaced input produces one stereotyped response, whose delay depends on the
	past only through the small residual of the previous excursion. Indeed, sparsely
	triggered excitable systems are
	contractive~\cite{cecconi2025contraction,bin2025reliability}, and contraction uniform
	over a compact set of inputs gives fading memory~\cite{matthews1993approximating}. In the
	one-to-one regime, they therefore admit an event operator~\eqref{eq:operator}, even
	though it cannot in general be computed in closed form. Its local gain around a periodic
	train can nevertheless be measured with the same experiment. Fig.~\ref{fig:event-gain}(Bottom) shows the result for a
	conductance-based neuron that fires by post-inhibitory rebound, forced through an
	inhibitory synaptic filter with different time constants. The neuron is entrained
	one-to-one above a minimal period, and beyond it the local gain decreases as the time
	between events increases. Therefore, seen at the events, a linear filter and an excitable
	neuron are not so different: the larger the time between events, the lower the gain.
	\section{Feedback interconnection of event operators}\label{sec:feedback}
	
	We now return to the feedback analysis in Section~\ref{sec:motivating}, exemplified by Fig.~\ref{fig:interconnection}, which we study from the viewpoint of event operators instead of continuous-time systems. The two systems are
	indexed by $i\in\{{\rm p},{\rm n}\}$, for pendulum and neuron, although the
	analysis does not depend on their physical nature. Each is modeled by an event
	operator $\Psi^i\colon(\cT^i)^{\Z}\to(\cD^i)^{\Z}$, with incremental gain $\gamma_i$
	and event describing function $\Delta^i(T)$. In feedback, the output events of each system
	become the input events of the other, so the input intervals are determined by the interconnection itself.
	
	In the one-to-one regime, the events alternate according to
	\[
	\cdots<t_{{\rm n},k}<t_{{\rm p},k}<t_{{\rm n},k+1}<t_{{\rm p}, k+1}<\cdots, \quad k \in \Z,
	\]
	where $t_{{\rm p},k}$ is the pendulum event triggered by the neuron event $t_{{\rm n},k}$. The
	delays are $\Delta^{\rm p}_k\coloneqq t_{{\rm p},k}-t_{{\rm n},k}$ and
	$\Delta^{\rm n}_k\coloneqq t_{{\rm n},k+1}-t_{{\rm p},k}$. Going around the loop
	from one input of a system to the next means traversing both delays once, so
	each interval is the sum of two consecutive delays
	\begin{equation}\label{eq:loop-identities}
		\begin{aligned}
			T^{\rm p}_k
			&=t_{{\rm n},k+1}-t_{{\rm n},k}
			=\Delta^{\rm p}_k+\Delta^{\rm n}_k,\\
			T^{\rm n}_k
			&=t_{{\rm p},k+1}-t_{{\rm p},k}
			=\Delta^{\rm n}_k+\Delta^{\rm p}_{k+1}.
		\end{aligned}
	\end{equation}
	
\begin{figure}[t]
		\centering
		\includegraphics[width=0.4\textwidth]{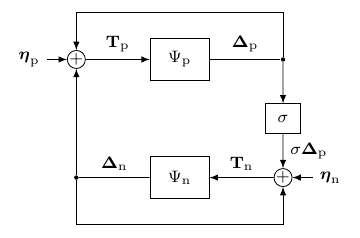}
		\caption{Block diagram of the loop equation~\eqref{eq:loop-equation}.
			Each delay sequence contributes to the input intervals of both
			subsystems, including its own through a self-loop.
			The sequences $\boldsymbol\eta^{\rm p}$ and
			$\boldsymbol\eta^{\rm n}$ represent timing perturbations.}
		\label{fig:feedback}
	\end{figure}

	A stability statement must say what happens when this timing is perturbed. We consider
	timing perturbations, such as finite transmission times, that enter the intervals
	additively, obtaining
	\begin{equation}\label{eq:loop-perturbed}
		\mathbf T^{\rm p}
		=\boldsymbol\Delta^{\rm p}
		+\boldsymbol\Delta^{\rm n}
		+\boldsymbol\eta^{\rm p},
		\quad
		\mathbf T^{\rm n}
		=\boldsymbol\Delta^{\rm n}
		+\sigma\boldsymbol\Delta^{\rm p}
		+\boldsymbol\eta^{\rm n},
	\end{equation}
	where $\boldsymbol\eta=(\boldsymbol\eta^{\rm p},\boldsymbol\eta^{\rm n})$. Each interval
	is now the sum of one delay from each system and a perturbation, and it must remain in
	the domain of its operator. Since the intervals of both systems take values in the same
	set, with $\cD^{\rm cl}\coloneqq\cD^{\rm p}+\cD^{\rm n}$ we can fix a perturbation bound
	$\bar\eta\ge0$ such that
	\begin{equation}\label{eq:admissible}
		\cD^{\rm cl}+[-\bar\eta,\bar\eta]
		\subseteq\cT^{\rm p}\cap\cT^{\rm n},
	\end{equation}
	that is, $\bar\eta$ is at most the distance from $\cD^{\rm cl}$ to the endpoints of
	$\cT^{\rm p}\cap\cT^{\rm n}$. The admissible perturbations are then
	\[\cW\coloneqq\set{\boldsymbol\eta\st
		\|\boldsymbol\eta^{\rm p}\|_\infty\le\bar\eta,\
		\|\boldsymbol\eta^{\rm n}\|_\infty\le\bar\eta}.
	\]
	Substituting \eqref{eq:loop-perturbed} into the event operators yields the
	\emph{loop equation}
	\begin{equation}\label{eq:loop-equation}
		\boldsymbol\Delta
		=\cF_{\boldsymbol\eta}(\boldsymbol\Delta)
		\coloneqq
		\begin{pmatrix}
			\Psi^{\rm p}(
			\boldsymbol\Delta^{\rm p}
			+\boldsymbol\Delta^{\rm n}
			+\boldsymbol\eta^{\rm p})\\
			\Psi^{\rm n}(
			\boldsymbol\Delta^{\rm n}
			+\sigma\boldsymbol\Delta^{\rm p}
			+\boldsymbol\eta^{\rm n})
		\end{pmatrix},
	\end{equation}
	illustrated in Fig.~\ref{fig:feedback}. The interconnection is now expressed
	entirely in terms of the two delay sequences.
	
	In order to show existence and stability of a fixed point of \eqref{eq:loop-equation}, we follow the classical small-gain
	argument~\cite{zames1966input,desoer2009feedback}, with two differences. The first is
	that our operators are defined only on admissible intervals, so we must check that
	the loop does not leave them. We work on the space $(\cD^{\rm p}\times\cD^{\rm n})^{\Z}$ of delay pairs, with the norm
	$\|\boldsymbol\Delta\|\coloneqq\|\boldsymbol\Delta^{\rm p}\|_\infty
	+\|\boldsymbol\Delta^{\rm n}\|_\infty$, and likewise for the perturbations.
	By~\eqref{eq:admissible}, for every $\boldsymbol\eta\in\cW$, $\cF_{\boldsymbol\eta}$ maps
	$(\cD^{\rm p}\times\cD^{\rm n})^{\Z}$ into itself. The second difference is
	the structure of the loop. In Fig.~\ref{fig:feedback}, each delay also enters the
	intervals of its own system, through a self-loop. To see what this implies, let
	$\boldsymbol\Delta,\tilde{\boldsymbol\Delta}\in(\cD^{\rm p}\times\cD^{\rm n})^{\Z}$ be two
	arbitrary delay pairs, and let $\mathbf T^i$ and $\tilde{\mathbf T}^i$ be the intervals
	they generate through~\eqref{eq:loop-perturbed}, with the same $\boldsymbol\eta\in\cW$. The
	perturbation cancels in the differences, which gives
	\begin{align*}
		\mathbf T^{\rm p}-\tilde{\mathbf T}^{\rm p}
		&=(\boldsymbol\Delta^{\rm p}-\tilde{\boldsymbol\Delta}^{\rm p})
		+(\boldsymbol\Delta^{\rm n}-\tilde{\boldsymbol\Delta}^{\rm n}),\\
		\mathbf T^{\rm n}-\tilde{\mathbf T}^{\rm n}
		&=(\boldsymbol\Delta^{\rm n}-\tilde{\boldsymbol\Delta}^{\rm n})
		+\sigma(\boldsymbol\Delta^{\rm p}-\tilde{\boldsymbol\Delta}^{\rm p}).
	\end{align*}
	Each operator therefore sees the difference of both delays. Since the shift does not
	change the supremum norm, the triangle inequality gives
	$\|\mathbf T^i-\tilde{\mathbf T}^i\|_\infty
	\le\|\boldsymbol\Delta-\tilde{\boldsymbol\Delta}\|$ for both systems. Combining this bound
	with the definition of the norm and of the incremental gains leads to
	\begin{equation}\label{eq:loop-gain-sum}
		\begin{aligned}
			&\|\cF_{\boldsymbol\eta}(\boldsymbol\Delta)
			-\cF_{\boldsymbol\eta}(\tilde{\boldsymbol\Delta})\|\\
			&\quad=\|\Psi^{\rm p}\mathbf T^{\rm p}-\Psi^{\rm p}\tilde{\mathbf T}^{\rm p}\|_\infty
			+\|\Psi^{\rm n}\mathbf T^{\rm n}-\Psi^{\rm n}\tilde{\mathbf T}^{\rm n}\|_\infty\\
			&\quad\le\gamma_{\rm p}\|\mathbf T^{\rm p}-\tilde{\mathbf T}^{\rm p}\|_\infty
			+\gamma_{\rm n}\|\mathbf T^{\rm n}-\tilde{\mathbf T}^{\rm n}\|_\infty\\
			&\quad\le(\gamma_{\rm p}+\gamma_{\rm n})
			\|\boldsymbol\Delta-\tilde{\boldsymbol\Delta}\|.
		\end{aligned}
	\end{equation}
	When $\gamma_{\rm p}+\gamma_{\rm n}<1$, $\cF_{\boldsymbol\eta}$ is a
	contraction, which gives our main result.
	
	\begin{theorem}[Event small gain]\label{thm:small-gain}
		Let $\Psi^{\rm p},\Psi^{\rm n}$ be event operators with incremental
		gains $\gamma_{\rm p},\gamma_{\rm n}$, let $\bar\eta\ge0$ satisfy
		\eqref{eq:admissible}, and suppose that
		\begin{equation}\label{eq:small-gain}
			\bar\gamma\coloneqq\gamma_{\rm p}+\gamma_{\rm n}<1.
		\end{equation}
		Then, for every $\boldsymbol\eta\in\cW$, the loop equation
		\eqref{eq:loop-equation} admits a unique solution
		$\cR(\boldsymbol\eta)\in(\cD^{\rm p}\times\cD^{\rm n})^{\Z}$.
		The closed-loop operator $\cR\colon\cW\to(\cD^{\rm p}\times\cD^{\rm n})^{\Z}$
		is strictly causal, commutes with the shift,
		$\cR\sigma=\sigma\cR$, and satisfies
		\begin{equation}\label{eq:loop-gain}
			\|\cR(\boldsymbol\eta)-\cR(\tilde{\boldsymbol\eta})\|
			\le
			\frac{\max\{\gamma_{\rm p},\gamma_{\rm n}\}}
			{1-\bar\gamma}
			\|\boldsymbol\eta-\tilde{\boldsymbol\eta}\|.
		\end{equation}
	\end{theorem}
	
	The theorem implies that the loop behaves as a single operator, from timing
	perturbations to delays: for each perturbation there is exactly one delay
	sequence, it depends causally on the perturbation, and~\eqref{eq:loop-gain}
	bounds the deviation that the perturbation produces. In particular, $\cR$ is
	Lipschitz, hence uniformly continuous, with respect to the topology induced by
	the supremum norm: a small change of the perturbation produces a small change of
	the delays, by the same bound everywhere in $\cW$. This is stability in the sense
	of classical operator theory~\cite{zames1966input,desoer2009feedback}, here in the
	event space.
	
	In the absence of external perturbations, namely $\boldsymbol\eta=0$, shift
	invariance forces the unique solution to be constant. Therefore, both systems are forced
	periodically with the same period, and the solution is determined by their periodic
	responses alone.
	
	\begin{corollary}[Periodic one-to-one rhythm]\label{cor:rhythm}
		Under the assumptions of Theorem~\ref{thm:small-gain},
		\[
		\cR(0)=
		\bigl(
		\Delta^{\rm p}(T^\star)\one,\,
		\Delta^{\rm n}(T^\star)\one
		\bigr),
		\]
		where $T^\star$ is the unique solution in
		$\cT^{\rm p}\cap\cT^{\rm n}$ of
		\begin{equation}\label{eq:balance}
			T^\star=\Phi(T^\star)
			\coloneqq\Delta^{\rm p}(T^\star)
			+\Delta^{\rm n}(T^\star).
		\end{equation}
		Moreover, for every $\boldsymbol\eta\in\cW$,
		\[
		\|\cR(\boldsymbol\eta)-\cR(0)\|
		\le
		\frac{\max\{\gamma_{\rm p},\gamma_{\rm n}\}}
		{1-\bar\gamma}\|\boldsymbol\eta\|.
		\]
	\end{corollary}
	
	The balance equation~\eqref{eq:balance}
	states that the period of the rhythm is the sum of the delays generated by that
	same period. Since $\Delta^{\rm p}$ and $\Delta^{\rm n}$ can be measured separately
	in open loop, the closed-loop period follows from a scalar equation, without
	simulating the interconnection. The same curves give a quick necessity test of the small-gain
	condition. Each $\Delta^i$ is $\gamma_i$-Lipschitz, so its slope
	$\lambda_i\coloneqq D\Delta^i(T^\star)$, when it exists, satisfies
	$|\lambda_i|\le\gamma_i$, and \eqref{eq:small-gain} cannot hold if
	$|\lambda_{\rm p}|+|\lambda_{\rm n}|\ge1$.
	
	\section{Simulation results}\label{sec:simulation}
	
\begin{figure}[!t]
		\centering
		\includegraphics[width=0.425\textwidth]{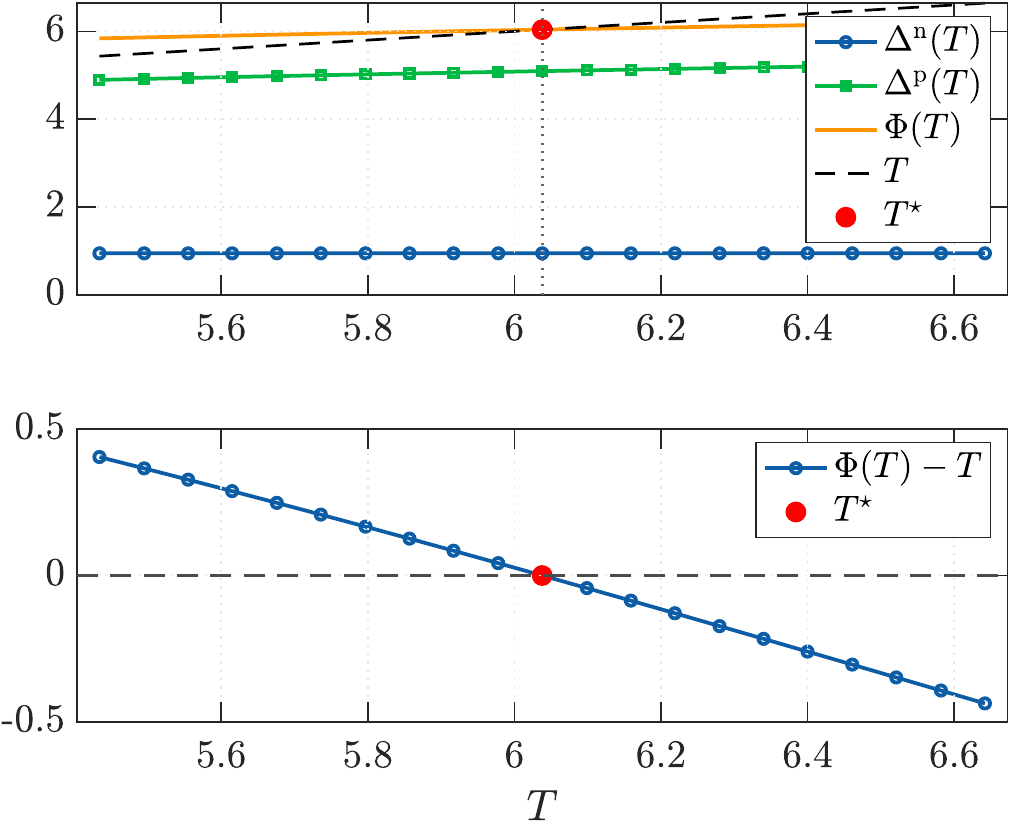}
		\caption{Balance condition of the neuron--pendulum loop. \emph{Top}:
			periodic responses $\Delta^{\rm n}(T)$ and $\Delta^{\rm p}(T)$, their
			sum $\Phi(T)$, and its intersection $T^\star$ with the diagonal; the
			dotted line marks the closed-loop period. \emph{Bottom}:
			$\Phi(T)-T$.}
		\label{fig:neuron_balance}
	\end{figure}

	We return to the neuron--pendulum loop of Fig.~\ref{fig:interconnection},
	with a conductance-based neuron that fires by post-inhibitory rebound, and
	each synapse modeled as the cascade of an event readout and a synaptic
	filter (Fig.~\ref{fig:readout-synapse}), as described in the Appendix.
	Fig.~\ref{fig:neuron_traj} shows the time behavior of the closed-loop solution: from rest, it settles within a few
	cycles onto a periodic one-to-one rhythm, in which each spike triggers one swing of
	the pendulum and each swing one spike. To predict this rhythm from the event
	description, we force each system periodically in open loop and record its periodic
	response. The two curves are shown in Fig.~\ref{fig:neuron_balance}. Their sum crosses
	the diagonal at $T^\star$, which coincides with the period of the closed loop, as
	predicted by the balance condition~\eqref{eq:balance}.
	
	\begin{figure}[ht]
		\centering
		\includegraphics[width=0.425\textwidth]{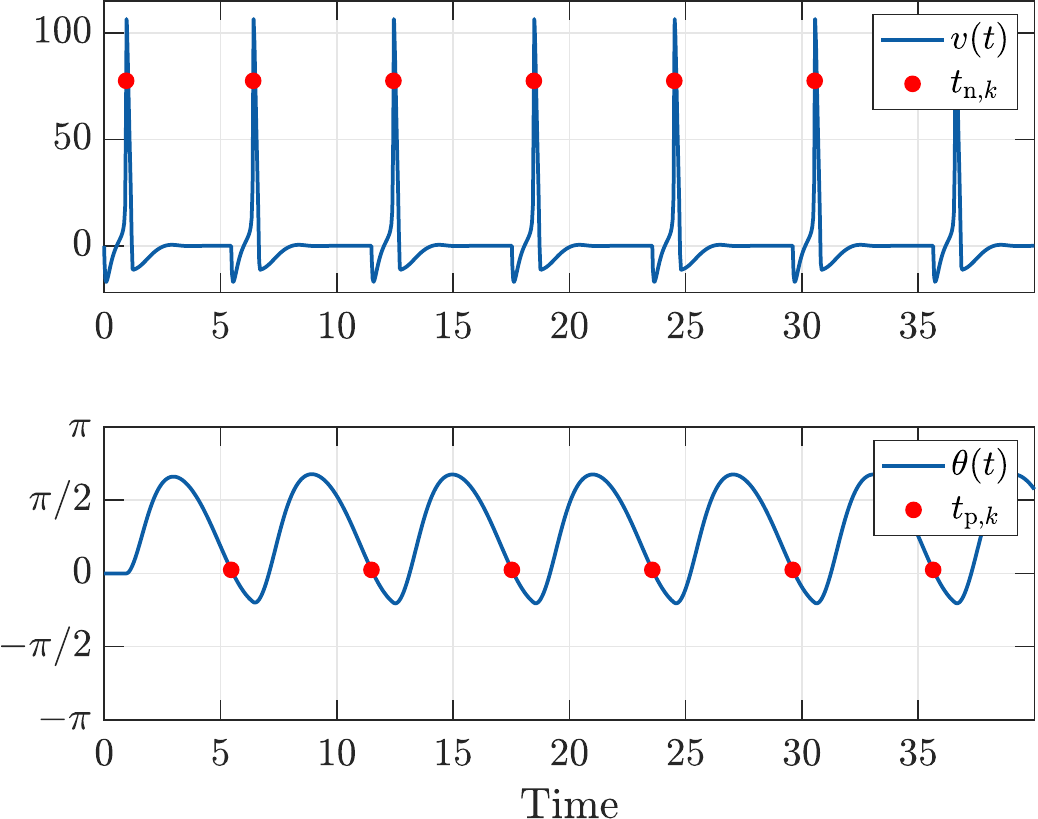}
		\caption{Closed-loop trajectories of the neuron--pendulum loop of
			Fig.~\ref{fig:interconnection}. \emph{Top}: membrane potential
			$v(t)$ and neuron events $t_{{\rm n},k}$. \emph{Bottom}: pendulum
			angle $\theta(t)$ and pendulum events $t_{{\rm p},k}$.}
		\label{fig:neuron_traj}
	\end{figure}
	
	Quantitatively, the balance gives $T^\star=6.039$, with
	$\Delta^{\rm n}(T^\star)=0.945$ and $\Delta^{\rm p}(T^\star)=5.094$. The slopes of the two curves at $T^\star$ are
	$\lambda_{\rm n}=1.1 \times 10^{-4}$ and $\lambda_{\rm p}=0.29$, so that
	$|\lambda_{\rm p}|+|\lambda_{\rm n}|=0.29<1$ and the necessary condition
	for~\eqref{eq:small-gain} is satisfied. The periodic response of the
	neuron is almost insensitive to variations of the period, consistent with an almost
	complete recovery between events.
	
	\section{Conclusions}\label{sec:conclusions}
	This paper reformulated the feedback analysis of neuromorphic systems
	in the event space. By modeling each synapse as the cascade of an event
	readout and a synaptic filter, the input-output behavior of each
	subsystem can be characterized by an operator mapping input event
	sequences to output event sequences. The stability of the underlying
	physical dynamics gives rise to fading-memory event operators, whose
	incremental gains can be shaped by imposing suitable sparsity
	conditions on the input event sequences. We established that a
	small-gain condition on the feedback interconnection guarantees
	the existence of a unique, stable solution to the feedback loop. In the absence of perturbations, this solution is a periodic
	one-to-one rhythm, with a period determined by a scalar balance
	equation that can be obtained by forcing the systems in open loop.
	
	The results of this paper suggest that the event-based nature of neuromorphic interactions naturally lends itself to a simpler and more tractable analysis when shifting the focus from continuous trajectories to the events.
	
	\section*{APPENDIX}
	
	\subsection{Event readout and synaptic filter}
	
	The simplest event readout is a threshold crossing. Given a signal $y$ and a threshold
	$y_{\rm th}$, we define the event section
	$\Sigma_y\coloneqq\set{(y,\dot y)\st y=y_{\rm th},\ \dot y>0}$, and the output events
	are the times at which $(y(t),\dot y(t))$ reaches $\Sigma_y$. The direction of the crossing
	is set by the condition on the velocity: replacing $\dot y>0$ with $\dot y<0$ gives
	downward crossings, as in the example of Section~\ref{sec:gain}. Every such crossing is
	transversal, so by the implicit function theorem the event time depends smoothly on
	the trajectory. This readout, however, has no refractoriness: the next upward crossing
	of the threshold produces a new event, however close it is to the previous one.
	Refractoriness is a crucial property of excitable systems, and we would like to impose it
	also on physical plants that do not necessarily exhibit it. To
	this end, we use a Schmitt-trigger-like readout~\cite[Ch.~3]{ChuaDesoerKuh1987}, described by
	\begin{equation}\label{eq:readout}
		\epsilon \dot q=-q+\tanh\bigl(\beta q+\alpha(y-y_{\rm th})\bigr),
	\end{equation}
	with gain $\alpha>0$, feedback strength $\beta>1$ and time constant $\epsilon>0$, for which $[-1,1]$ is forward invariant. At fixed
	$y$, \eqref{eq:readout} is bistable for $y\in(y_{\lo},y_{\hi})$, with
	$y_{\hi},y_{\lo}\coloneqq y_{\rm th}\pm\omega_\beta/\alpha$ and
	$\omega_\beta\coloneqq\sqrt{\beta^2-\beta}-\operatorname{arccosh}\sqrt\beta>0$. For small
	$\epsilon$, the readout thus switches up only when $y$ exceeds $y_{\hi}$, and it must
	switch down, which requires $y$ to fall below $y_{\lo}$, before it can switch up
	again. This hysteresis is the refractoriness of the readout.
	Events are now defined on the state of the readout, through the event section
	$\Sigma_q\coloneqq\set{(q,y)\in[-1,1]\times\mathbb R\st q=0,\ y>y_{\rm th}}$, as the
	times $\mathcal I=\{t_k\}\coloneqq\set{t\st(q(t),y(t))\in\Sigma_q}$. On $\Sigma_q$,
	\eqref{eq:readout} gives $\epsilon\dot q=\tanh(\alpha(y-y_{\rm th}))>0$, so every
	crossing is transversal and upwards, without any condition on $\dot y$, and the event
	time again depends smoothly on the trajectory. Conversely, $\dot q\le0$ at $q=0$ when
	$y\le y_{\rm th}$, so that $q$ cannot cross zero upwards twice unless $y$ has reached
	$y_{\rm th}$ from above in between.
	
	The synaptic filter maps the sequence of events
	$\mathcal I$ generated by the readout into a
	conductance signal. It evolves
	according to the impulsive dynamics
	\begin{equation}\label{eq:synaptic-filter}
		\begin{aligned}
			\dot s = f_s(s), && t\notin\mathcal I,\qquad
			s(t^+) = \kappa(s(t^-)), && t\in\mathcal I,
		\end{aligned}
	\end{equation}
	with output $g(t)=h_s(s(t))\geq 0$, where $f_s$, $\kappa$ and $h_s$ are smooth.
	The
	state evolves on a compact set $\mathcal S\subset\mathbb R^{n}$ that is forward
	invariant under both the continuous flow and the reset map. Consequently, for every locally finite
	sequence of input events and every initial condition in $\mathcal S$, the solution
	of~\eqref{eq:synaptic-filter} is forward complete and remains in $\mathcal S$. In
	particular, $g$ is uniformly bounded.
	
	\subsection{Proof of Proposition~\ref{prop:fading}}
	Let $w$ be a memory kernel, $L\coloneqq T_{\max}-T_{\min}$, and fix
	$\varepsilon>0$. For all $v,\tilde v\in\cT^{\mathbb N}$ we have
	$|v_j-\tilde v_j|\le L$, so the series in~\eqref{eq:fading} is dominated
	by $L\sum_{j\ge1}\rho_j<\infty$. Its tails therefore vanish, and we fix
	$J\in\mathbb N$ such that $L\sum_{j>J}\rho_j<\varepsilon/2$. For $j\le J$,
	we have $w_j|v_j-\tilde v_j|\le\|v-\tilde v\|_w$ and, by monotonicity of
	the kernel, $w_j\ge w_J>0$, so that
	$|v_j-\tilde v_j|\le\|v-\tilde v\|_w/w_J$. Splitting the sum
	in~\eqref{eq:fading} at $J$ and letting $S_J\coloneqq\sum_{j\le J}\rho_j$
	gives
	\[
	|\psi(v)-\psi(\tilde v)|<\frac{S_J}{w_J}\|v-\tilde v\|_w+\frac{\varepsilon}{2}.
	\]
	Let $\delta\coloneqq\varepsilon w_J/(2(S_J+1))$. If
	$\|v-\tilde v\|_w<\delta$, the first term is at most
	$\varepsilon S_J/(2(S_J+1))<\varepsilon/2$, hence
	$|\psi(v)-\psi(\tilde v)|<\varepsilon$. Since $\delta$ does not depend on $v$ and $\tilde v$, $\psi$ is uniformly continuous with respect to $\|\cdot\|_w$.
	\hfill$\blacksquare$
	
	\subsection{Event operator of the first-order filter}
	For every $v\in\cT^{\mathbb N}$, the ages satisfy
	$jT_{\min}\le a_j\le jT_{\max}$, so~\eqref{eq:lag-functional}
	and~\eqref{eq:lag-periodic} give
	$\Delta(T_{\max})\le\psi(v)\le\Delta(T_{\min})$. The delays thus lie in
	$\cD=[\Delta(T_{\max}),\Delta(T_{\min})]$, with
	$\Delta(T_{\max})\ge\tau\ln(b/y_{\rm th})>0$, and the remaining requirement
	$\Delta(T_{\min})<T_{\min}$ is equivalent to
	$T_{\min}>\tau\ln(1+b/y_{\rm th})$.
	
	Let $p\coloneqq e^{-T_{\min}/\tau}$ and $W_j\coloneqq\sum_{i\ge j}e^{-a_i/\tau}$.
	Differentiating~\eqref{eq:lag-functional} gives $\partial\psi/\partial v_j=-W_j/W_0$.
	Since $a_{i+1}\ge a_i+T_{\min}$, we have $W_{j+1}\le pW_j$, hence
	$|\partial\psi/\partial v_j|\le p^j$ on $\cT^{\mathbb N}$. Integrating along the segment
	between $v$ and $\tilde v$, which lies in the convex set $\cT^{\mathbb N}$,
	gives~\eqref{eq:fading} with $\rho_j=p^j$, so $\gamma\le\sum_{j\ge1}p^j=p/(1-p)$. This
	value is attained, since $\Delta$ is $\gamma$-Lipschitz and
	$|D\Delta(T_{\min})|=p/(1-p)$, so that $\gamma=p/(1-p)=\gamma_{\rm loc}(T_{\min})$.
	\hfill$\blacksquare$
	
	\subsection{Proof of Theorem~\ref{thm:small-gain}}
	\emph{Existence and uniqueness.} Let $\boldsymbol\eta\in\cW$ and
	$\mathcal X\coloneqq(\cD^{\rm p}\times\cD^{\rm n})^{\Z}$. For
	$\boldsymbol\Delta\in\mathcal X$, each interval in~\eqref{eq:loop-perturbed}
	lies in $\cT^{\rm p}\cap\cT^{\rm n}$ by~\eqref{eq:admissible}, so both operators
	can be applied and $\cF_{\boldsymbol\eta}$ maps $\mathcal X$ into itself. The set
	$\mathcal X$ is closed in $\ell^\infty(\Z)\times\ell^\infty(\Z)$, hence complete,
	and $\cF_{\boldsymbol\eta}$ is a contraction on it by~\eqref{eq:loop-gain-sum}
	and~\eqref{eq:small-gain}. The Banach fixed-point theorem
	(see~\cite[Ch.~1]{bullo_contraction_2024}) gives a unique fixed point
	$\cR(\boldsymbol\eta)\in\mathcal X$.
	
	\emph{Shift invariance.} The shift maps $\cW$ and $\mathcal X$ into themselves and
	commutes with $\Psi^{\rm p}$ and $\Psi^{\rm n}$, so
	$\cF_{\sigma\boldsymbol\eta}(\sigma\boldsymbol\Delta)=\sigma\cF_{\boldsymbol\eta}(\boldsymbol\Delta)$.
	Hence $\sigma\cR(\boldsymbol\eta)$ is a fixed point of $\cF_{\sigma\boldsymbol\eta}$,
	and uniqueness gives $\sigma\cR(\boldsymbol\eta)=\cR(\sigma\boldsymbol\eta)$.
	
	\emph{Strict causality.} Fix $K\in\Z$ and let $\pi_K$ restrict sequences to the
	indices $k\le K$. By strict causality of $\Psi^{\rm p},\Psi^{\rm n}$
	and~\eqref{eq:loop-perturbed}, $\pi_K\cF_{\boldsymbol\eta}(\boldsymbol\Delta)$
	depends only on $\pi_K\boldsymbol\Delta$ and on $\eta^i_j$ with $j<K$, so
	$\cF^K_{\boldsymbol\eta}(\pi_K\boldsymbol\Delta)\coloneqq\pi_K\cF_{\boldsymbol\eta}(\boldsymbol\Delta)$
	is well defined on $\pi_K\mathcal X$, which is complete with the norm restricted to
	$k\le K$. Extending $x,\tilde x\in\pi_K\mathcal X$ with a common tail gives
	$\boldsymbol\Delta,\tilde{\boldsymbol\Delta}\in\mathcal X$ with
	$\|\boldsymbol\Delta-\tilde{\boldsymbol\Delta}\|=\|x-\tilde x\|$,
	so~\eqref{eq:loop-gain-sum} makes $\cF^K_{\boldsymbol\eta}$ a contraction. Applying
	$\pi_K$ to $\cR(\boldsymbol\eta)=\cF_{\boldsymbol\eta}(\cR(\boldsymbol\eta))$ shows
	that its unique fixed point is $\pi_K\cR(\boldsymbol\eta)$. If
	$\eta^i_j=\tilde\eta^i_j$ for all $j<K$, then
	$\cF^K_{\boldsymbol\eta}=\cF^K_{\tilde{\boldsymbol\eta}}$, hence
	$\pi_K\cR(\boldsymbol\eta)=\pi_K\cR(\tilde{\boldsymbol\eta})$.
	
	\emph{Gain.} Let $\boldsymbol\Delta=\cR(\boldsymbol\eta)$ and
	$\tilde{\boldsymbol\Delta}=\cR(\tilde{\boldsymbol\eta})$. Since these are fixed points,
	the triangle inequality gives
	\[
	\|\boldsymbol\Delta-\tilde{\boldsymbol\Delta}\|
	\le\|\cF_{\boldsymbol\eta}(\boldsymbol\Delta)-\cF_{\boldsymbol\eta}(\tilde{\boldsymbol\Delta})\|
	+\|\cF_{\boldsymbol\eta}(\tilde{\boldsymbol\Delta})-\cF_{\tilde{\boldsymbol\eta}}(\tilde{\boldsymbol\Delta})\|.
	\]
	The first term is at most $\bar\gamma\|\boldsymbol\Delta-\tilde{\boldsymbol\Delta}\|$
	by~\eqref{eq:loop-gain-sum}. In the second, the inputs of $\Psi^i$ differ by
	$\boldsymbol\eta^i-\tilde{\boldsymbol\eta}^i$, so it is at most
	\[\gamma_{\rm p}\|\boldsymbol\eta^{\rm p}-\tilde{\boldsymbol\eta}^{\rm p}\|_\infty
	+\gamma_{\rm n}\|\boldsymbol\eta^{\rm n}-\tilde{\boldsymbol\eta}^{\rm n}\|_\infty
	\le\max\{\gamma_{\rm p},\gamma_{\rm n}\}\|\boldsymbol\eta-\tilde{\boldsymbol\eta}\|.\]
	Since $\bar\gamma<1$, rearranging gives~\eqref{eq:loop-gain}.
	\hfill$\blacksquare$
	
	\subsection{Proof of Corollary~\ref{cor:rhythm}}
	
	Since $\sigma0=0$, Theorem~\ref{thm:small-gain} gives $\sigma\cR(0)=\cR(0)$, so $\cR(0)$ is
	constant, that is, $\cR(0)=(\Delta^{\rm p}_\star\one,\Delta^{\rm n}_\star\one)$ for some
	$\Delta^i_\star\in\cD^i$. By~\eqref{eq:loop-perturbed} with $\boldsymbol\eta=0$, both
	systems then receive the constant intervals $T^\star\one$, where
	$T^\star\coloneqq\Delta^{\rm p}_\star+\Delta^{\rm n}_\star\in\cD^{\rm cl}\subseteq\cT^{\rm p}\cap\cT^{\rm n}$.
	Hence $\Delta^i_\star\one=\Psi^i(T^\star\one)=\Delta^i(T^\star)\one$, and summing the two
	delays gives~\eqref{eq:balance}.
	
	For uniqueness, let $T\in\cT^{\rm p}\cap\cT^{\rm n}$ solve~\eqref{eq:balance}. Then
	$(\Delta^{\rm p}(T)\one,\Delta^{\rm n}(T)\one)\in\mathcal X$ solves the loop equation with
	$\boldsymbol\eta=0$, so it equals $\cR(0)$ by the uniqueness in
	Theorem~\ref{thm:small-gain}, and $T=T^\star$. Finally, the bound follows
	from~\eqref{eq:loop-gain} with $\tilde{\boldsymbol\eta}=0$.
	\hfill$\blacksquare$


	\bibliographystyle{IEEEtran}
	\bibliography{main_final}
	
\end{document}